\documentclass[10pt]{iopart}

\usepackage{graphicx}
\usepackage{dcolumn}
\usepackage{bm}
\usepackage{epstopdf}
\usepackage{iopams}

\begin{document}

\title[Measure the pseudo-Stark shift by the photon echo beating]
{Stark shift in Y$_2$SiO$_5$:Er$^{3+}$ by the photon echo beating method}

\author{V. N. Lisin$^{1}$, A. M. Shegeda$^{1}$, V. V. Samartsev$^{1}$, S.A. Kutovoi$^{1,2}$, Yu.D. Zavartsev$^{1,2}$}

\address{{$^{1}$}Zavoisky Physical-Technical Institute, FRC "Kazan Scientific Center of RAS", Sibirsky tract, 10/7, Kazan, 420029, Russia}
\address{{$^{2}$}Prokhorov General Physics Institute, Russian Academy of Sciences, Vavilova St., 38, Moscow, 119991, Russia}

\ead{valerylisin@gmail.com}
\vspace{10pt}
\begin{indented}
\item[]Juli 2018
\end{indented}

\begin{abstract}
The coefficient of a linear Stark shift of the $^4$F$_{9/2}$ - $^4$I$_{15/2}$ transition of Er$^{3+}$ ion in Y$_2$SiO$_5$ is measured (14.3 kHzV$^{-1}$cm ) by splitting the upper and lower levels of the transition by a pulsed electric field and measuring the splitting frequency from the period of beatings of the temporal form of a photon echo signal (stark photon echo beating method). The linear Stark coefficients of Er$^{3+}$ transitions in Y$_2$SiO$_5$ have not been measured before, and this is a valuable result for a number of applications such as gradient echo memory, group velocity control via spectral hole burning, etc.

\end{abstract}



%

\ioptwocol

\section{Introduction}

This paperТs purpose is to observe the photon echo beating (PEB) of the $^4$F$_{9/2}$ - $^4$I$_{15/2}$ line of Er$^{3+}$ ion in Y$_2$SiO$_5$ by applying weak electric field pulses and to measure the pseudo-Stark splitting the optical line from the period of beats of the temporal form of the photon echo signal. The term "photon echo" (PE) refers to coherent radiation from a medium in the form of a short pulse, caused by restoration of phase of separate radiators after the change of sign of their relative frequency. If perturbation splits optical line, for example, on two lines, this means that the frequency of transitions in two groups of radiators are shifted by different values. If a pulse of the perturbation overlaps in time with the echo-pulse, then the echo waveform changes and photon echo beats arise. The PEB have been observed initially in systems in which Zeeman splitting is realized (ZPEB) [1$-$3]. In systems in which the pseudo-Stark effect is manifest, the line splitting in electric field is due to the different Stark shifts at different lattice sites, for example A and B sites. Transformation of the A site into the B site is only possible by symmetry operations involving inversion, while transformation of A to A or B to B is achieved solely by translation or by both translation and rotation around the optic axis. The A and B sites are energetically equivalent in the absence of an electric field. Beats of the time shape of the PE signal in a pulsed electric field (SPEB) were first observed by us in [4, 5]. This made it possible to determine with good accuracy the magnitude of the pseudo-Stark splitting of chromium ions in ruby by the SPEB method.
We could not find the work of other authors, where the numerical value of the Stark coefficient for the Er$^{3+}$ ion in Y$_2$SiO$_5$ was determined. In the infrared range, a PE at a wavelength of 1.5 $\mu$m was observed in [6] but the numerical value of the $\partial\nu/\partial E$ was not determined.

\section{Y$_2$SiO$_5$ in electric field}

Y$_2$SiO$_5$ (YSO) single crystals are monoclinic with the C$^6_{2h}$ = I2/a space group. [7]. A unit cell comprises eight molecules of YSO. Yttrium atoms occupy two crystallographic sites (Y$_1$ and Y$_2$) with a triclinic local symmetry C$_1$. The sites Y$_1$ and Y$_2$ are distinguished by their coordinate numbers of six and seven respectively, which indicates the number of bonds each site has to surrounding oxygen atoms. Silicon and oxygen ions also occupy sites with the local symmetry C$_1$. Figure 1 shows the YSO structure in a projection to the $ac$ plane.

 \begin{figure}

\includegraphics[scale=0.5,viewport=80 400 404 800]{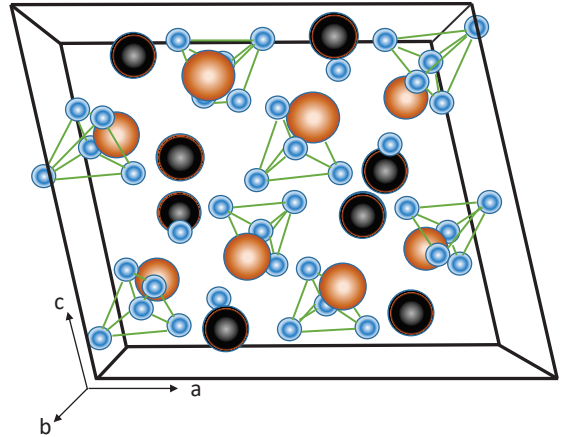}
\caption{\label{fig:stark} Crystal structure of Y$_2$Si0$_5$. Shown is a single unit cell. The data for the
positions of the atoms are taken from Maksimov et al. [7]. The silicon atoms form the centers of
the tetragons with oxygens at the ends. The small spheres are the remaining 8 oxygens and the
large spheres represent the 16 Yttria. The two inequivalent sites of Yttrium are colored white and
black. }
\end{figure}

 Because the symmetry group I2/a of the YSO lattice has a center of symmetry and the local symmetry of $C_1$ is not, the eight Y$_j$ (j= 1, 2) ions in the unit cell, can be broken up into pairs of 2A1$_j$, 2B1$_j$, 2A2$_j$, 2B2$_j$ ions, dipole moments A$_{ij}$ and B$_{ij}$ have opposite directions (i=1, 2). The dipole moments of the A1$_j$ and A2$_j$ ions are arranged symmetrically about the $b$ axis at an angle $\theta $ to the $b$ axis. For Pr$^{3+}$:YSO this angle is $\theta $ = 12.4$^o$ [8]. Erbium ions replace Y ions in the unit cell. The shifts $\delta\nu$ of transition frequencies of A$_i$ and B$_i$ ions in the electric field $E$ equal in magnitude but differ in sign and the optical line splits:

\begin{eqnarray}\label{eq1}
&\delta \nu=\pm \frac{\delta \mu L}{h} \cos{(\theta)}E=\partial \nu/\partial E\cdot E,\nonumber\\
&\frac{\delta \mu L}{h}=\frac{\partial \nu / \partial E}{\cos(\theta)},   L=\frac{\varepsilon+2}{3}\approx\frac{n^2+2}{3}.
\end{eqnarray}

Here the electric field $E$ is directed along the $b$ axis, $\delta \mu$ is the difference between the electric dipole moments of the states between which the transition is observed, $L$ is the Lorentz correction, $\varepsilon$ is the low-frequency dielectric constant of the sample, which for molecular crystals can be expressed in terms of the square of the refractive index $n$, $\theta$ is the angle between the  ${\delta \mu}$ and $E$ directions.

\section{Photon echo beating }

If the electric field pulse (EP) is turned on, for example, after the second laser pulse, then the total dipole moment from each pair of oppositely directed dipoles A$_{ij}$ and B$_{ij}$ is defined as

\begin{eqnarray}
\label{eq2}
P(t)=P_0(t)\cos{(2\pi\int^t_{t_0}\delta \nu(t\prime)dt\prime)}= \nonumber\\
P_0(t)\cos(2\pi\frac{\delta \mu L}{h}\cos(\theta)\int^t_{t_0}E(t\prime)dt\prime).
\end{eqnarray}

Here, $P_0(t)$ and $t_0$ are the dipole moment in the absence of EP and the switching-on time of EP, respectively. Since the dipole moments from the other pair are directed at the same angle $\theta$, the expression for the resulting dipole moment from all the ions Y1 has the same form (2).
In the absence of a dc external electric field $E_0$, it can be shown similarly to [9] that, independently of the coherence of the exciting laser, the expression for the intensity of the echo $I$ can be written in the following form:

\begin{eqnarray}
\label{eq3}
  I(t)=I_0(t)(1+\cos(4\pi\int^t_{t_0}\delta \nu(t^\prime)dt^\prime))/2=\nonumber\\
  =I_0(t)(1+\cos(4\pi \frac{\delta \mu L}{h} \cos{(\theta)}\int^t_{t_0}E(t^\prime)dt^\prime))/2
\end{eqnarray}
where $I_0(t)$ is intensity of an echo in the absence of an electric pulse. When deriving expressions (2) and (3), it was taken into account that the Photon Echo intensity is proportional to the square of the electric dipole moment module. The electric dipole moment is equal to the spur of the product of the electric dipole moment operator on the off-diagonal terms of the density matrix. Given that the polarization of the laser light is directed along the $b$ axis, as well as the EP field, the optical transitions are carried out between the States with the same projection of the electron spin in the ground and excited States. This means that the off-diagonal elements of the density matrix are diagonal in electron spin projections.

Figure 2, for example, shows oscillograms of the observed signals in YSO for different values of the pulsed electric field.
The pulse is applied simultaneously with the echo signal. You can see the appearance of beats of the photon echo and the change in their frequency with increasing $E$.

\begin{figure}
\includegraphics[scale=0.45,viewport=80 110 204 550]{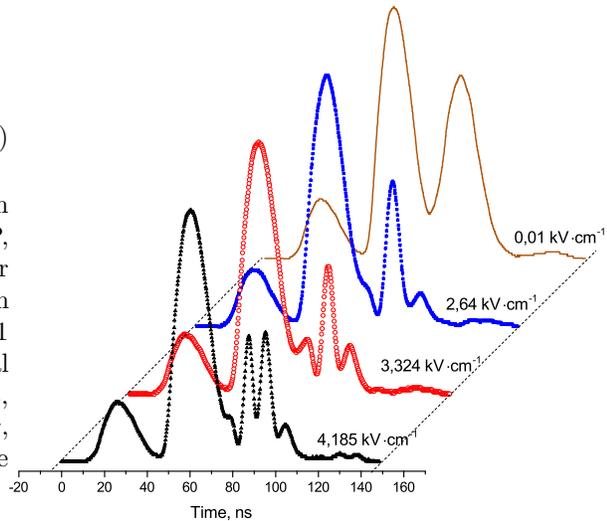}
\caption{\label{fig:beats}
Oscillograms of the observed signals in the Y$_2$SiO$_5$:Er$^{3+}$ sample for different values of the electric field. The value of the electric field correspond to the maximum of the  EP amplitude. The  right pulse is the echo signal. In its form, beats are observed, whose frequency increases with increasing an electric field. The 10V/cm pulse does not change the shape of the echo signal}
\end{figure}

The electric field can be expressed in terms of the potential difference $V$ and the sample thickness $d$:
\begin{equation}
\label{eq4}
  E(t)=V(t)/d
\end{equation}

\section{Experimental conditions}
Erbium doped Y$_{2}$SiO$_{5}$ crystals were grown by the Czochralski method in Ir crucibles in the 99 vol.\% Ar + 1 vol.\% O$_{2}$ atmosphere. The purity of primary components, i.e. Y$_{2}$O$_{5}$ and Nd$_{2}$O$_{3}$ was no worse than 99.75\%. We studied theY$_{2}$SiO$_{5}$  monocrystals doped with 0.005 at.\% $^{}$Er$^{3+}$.

The experiment was carried out at a temperature of 2K. Using a tunable dye laser Oksazin 17, the transmission spectrum of our sample with an erbium concentration of 0.005 has been mesured. The laser light propagated at an angle of 35 degrees to the $D_1$ axis, the polarization of the electric field of the laser is parallel to $b$ axis. (see figure 3).

\begin{figure}
\includegraphics[scale=0.6,viewport=120 420 404 750]{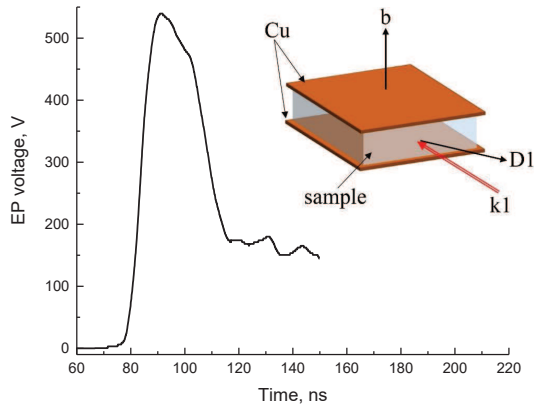}
\caption{\label{fig:geom} The geometry of the experiment and the oscillogram of the voltage pulse on the copper plates of the capacitor. A sample of Y$_2$SiO$_5$:Er$^{3+}$ with a thickness of 1.3 mm is in the capacitor. The angle between the direction of the laser pulse k$_1$ and the D$_1$ axis of the crystal is 35$^o$. The axis $b$ is perpendicular to $k_1$. The electric field laser $E_L$ is parallel to the axis $b$ as well as the electric field of the capacitor}
\end{figure}

The two lower frequency transitions  corresponded to wavelengths of 659 nm and 657.87 nm and relate to two nonequivalent erbium positions in the lattice Y$_1$ and Y$_2$ accordingly [10]. We can be to observe the photon echo only at a wavelength of 657.87 nm (Y$_2$ site). Absorption at 659.1 nm is small and PE has not been observed at this transition. The PE signal was recorded  in the reversed mode with a delay time between pulses $t_{12}$ $\sim$ 33 ns (when the delay time between laser pulses $t_{12}$ increased, the echo intensity was small to control the temporal shape of the PE). The recorded signal from a high-speed photodetector was directed to a digital oscilloscope Tektronix TDS 2022 oscilloscope with the number of accumulations equal to 64. Measurements and data processing took place in the LabView environment.
To study the effect of a pulsed electric field on the photon echo, we cut out a sample of thickness $d$ = 1.30$\pm$0.01 mm perpendicular to the $b$ axis. The sample was placed in a capacitor of two copper plates 0.1 mm thick and tightly pressed to the faces perpendicular to the $b$ axis (the electric field created by the capacitor is parallel to the $b$ axis).  It should be noted that the rectangular shape pulse might fit better purpose of our measurements. Preliminary experiments have shown that the amplitude of the electric field generated by the generator of rectangular pulses, which we used in [9] to measure the linear Stark shift coefficient of Cr3+ ion, is insufficient for the occurrence of beatings in Y$_2$SiO$_5$:Er$^{3+}$. Therefore, a generator was built on avalanche transistors according to the scheme Arkadiev Ц Marx, which makes it possible at a voltage of 300 V to obtain nanosecond pulses with an amplitude up to 544 V with a short leading edge and a shape typical of a capacitive relaxator. The amplitude of the pulses was varied by a step attenuator in steps of 1 dB.

The difference between the EP shape and the rectangular shape is not of fundamental importance,because the beats of the PE are determined by the area of an electric field pulse (see (3)), and the area of a pulse is automatically determined in the experiment. The switching-on time of EP was controlled by a programmable delay in a wide range in steps of 0.1 ns. The duration of such a pulse is longer than the delay time $t_{12}$ between laser pulses. This makes it difficult to use the oscillations of the intensity of the PE to determine the Stark shift of the line. When using the PEB method, there are no restrictions related to the duration of the EP. Only a part of the EP area, which overlaps with the echo pulse in time, is responsible for the results.

However, it is important to know exactly the shape of the EP, causing the beeping of the photon echo signal. It turned out that the shape of the trailing edge of the pulses of the Arkadiev--Marx generator depends on the load. Therefore, for each position of the step attenuator, which is the load of the generator, it was necessary to register the exact shape of the pulses for a correct comparison of the theory and experiment.

\section{Results}
Figure 2 shows the change in the shape of the PE signal when the electric pulse is turned on. A detailed consideration of the shape of the echo is shown in figure 4.

\begin{figure}
\includegraphics[width= 8.5cm]{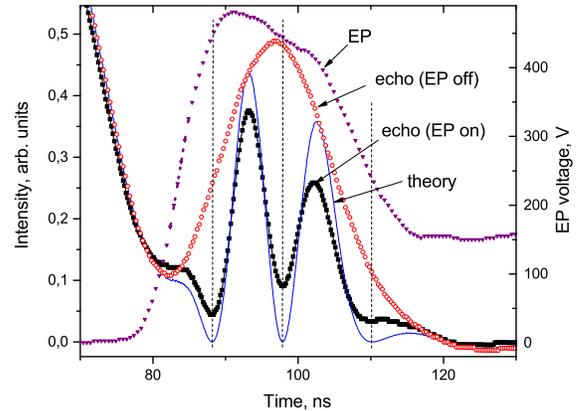}
\caption{\label{fig:geom} Beats of the shape of the PE signal when the pulse is turned on 485 V. Squares show an experiment, a solid line is a theory. The circles show the observed signal without EP}
\end{figure}

When recording the beats of the PE in the experiment, the time of appearance of the first minimum in the time form of the PE depends on the switching-on time  of the EP and its amplitude. For each amplitude E, the turn-on time $\it t_0$ was selected in such a way that the minima were clearly visible.
Equation (3) describes the beats of the temporal form of the PE vs on the Stark coefficient, the amplitude of the EP, and its switching-on time.
The parameter $\partial\nu/\partial E$ was found from the condition that the points of the first and second minimum of the echo form (see figure 4) coincide with those obtained theoretically with the help of (3). For each value of the pulsed electric field, the value of the $\partial\nu/\partial E$  was determined, at which coincidence with the experiment was the best. Deviation of the $\partial\nu/\partial E$  from this value more than 1 $\%$ caused a noticeable discrepancy between theory and experiment.

The parameter $\partial\nu/\partial E$ vs $E=V/d$ is shown in figure~5. You can see in figure~5, that
\begin{equation}
\label{eq5}
  \partial\nu/\partial E=14.3\pm0.17~kHzV^{-1}cm
\end{equation}
Here we take into account that the root-mean-square error of the values of $\partial\nu/\partial E$, given in figure 5, is 0.01~kHzV$^{-1}$cm. Let's take into account that the error of measurement at each point is approximately equal to 1 percent: 0.143~kHzV$^{-1}$cm. Adding the error squares and extracting the root, we obtain for the resulting error a value of 0.17~kHzV$^{-1}$cm.
We also take into account the fact that $d$ = 0.130$\pm$0.001~cm.

Exactly the same result can be obtained if one simlply calculates the area of the EP isolated in figure~4 by dashed lines between the two nearest minima. Equating the argument value under the cosine sign in (3) to $2\pi$, we get the value of the required parameter $\partial\nu/\partial E$.

In fact, the above error reflects only the accuracy of the comparison of theory and experiment. However, there are reasons that we have not taken into account, which reduce the accuracy of measurements, for example, a small jitter between the laser and electric pulses, an error of 2\% of the probe, by means of which the amplitude of the pulses was measured. Therefore, it will be more accurate to evaluate the result obtained in the form:
\begin{equation}
\label{eq6}
  \partial\nu/\partial E=14.3\pm0.7~kHzV^{-1}cm
\end{equation}

\begin{figure}
\includegraphics[width= 8.5cm]{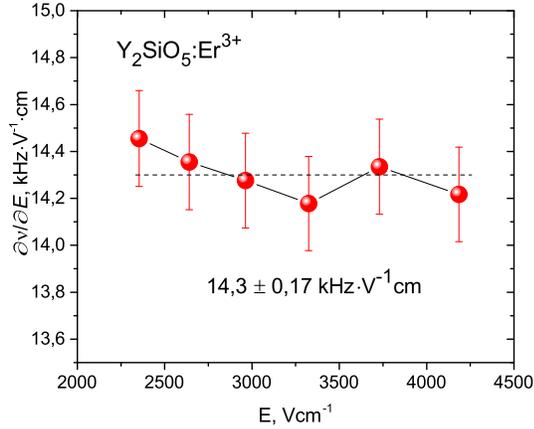}
\caption{\label{fig:stark}  Stark-shift parameter $\partial\nu/\partial E$ vs $E=V/d$}
\end{figure}

\section{Discussion}
The Stark-shift coefficient  $\partial\nu/\partial E = 17.7 \pm 0.3$~kHzV$^{-1}$cm for the Er$^{3+}$ ion in the YAlO$_3$ matrix for the same $^4$F$_{9/2}$ - $^4$I$_{15/2}$ transition was found in [11].
The effective dipole moment  $\delta \mu L/h= 15 \pm 1$~kHzV$^{-1}$cm  for Er$^{3+}$ in SiO${_2}$ has been found in [12] from an increase in the transmission width of the spectral hole with an increase in the external electric field.
As follows from (6) and (1)  the value of this parameter for Er$^{3+}$:Y${_2}$SiO$_{5}$ is $\delta \mu L/h =14.3 \pm 0.7$~kHzV$^{-1}$cm$/\cos(\theta) > 14.3 \pm 0.7~$kHzV$^{-1}$cm.
It can be seen that a values of the Stark-shift coefficients for the f-ion (Er$^{3+}$) depends weakly on the matrix. The Stark shift of the d-ion (Cr$^{3+}$: $\partial\nu/\partial E$ = 107~kHz$^{-1}$cm [5]) is much (6 times) larger than the Stark shift of f-ion (Er$^{3+}$).

\section{Conclusion}
We again applied the method of beats of the photon echo (PEB) and, thanks to this, we for the first time determined the Stark shift of the erbium ion in YSO. This result shows that it is possible to determine the Stak shift with good accuracy from a single oscillogram if we learn the parameters of the EP that caused the PE beating. For a number of applications related to quantum memory, it would be important to know the Stark shift coefficient at the 1.5~$\mu m$  transition of Er. The stark photon echo beating method also can be used in this case. We plan to do this in the near future.

\section{Acknowledgements}
This work is partially supported by the Russian Science Foundation (project no. 16-12-00041).



\Bibliography{8}
 \bibitem{1}  Lisin V, Shegeda M 2012 Modulation of the Shape of the Photon
Echo Pulse by a Pulsed Magnetic Field: Zeeman splitting in
LiLuF4:Er3 + and LiYF4:Er3 + \textit{JETP Letters} \textbf{96}
328-332

\bibitem{2}  Lisin V, Shegeda M, Samartsev V 2015 The application of the weak
magnetic field pulse to measure g-factors of ground and excited optical
states by a photon echo method  \textit{Laser Phys. Lett.} \textbf{12} 025701

\bibitem{3}  Lisin V, Shegeda M, Samartsev V 2015 New possibilities of photon echo:
determination of ground and excited states g-factors applying a weak magnetic
field pulse  \textit{Journal of Physics: Conference Series} \textbf{613} 012013

\bibitem{4} Lisin V, Shegeda M and Samartsev V 2015 Definition of shifts of optical transitions frequencies due to pulse perturbation action by the photon echo signal form  \textit{EPJ Web Conf. } \textbf{103} 07004

\bibitem{5}Lisin V, Shegeda A, Samartsev V 2016 The application of weak electric field pulses to measure the pseudo-Stark split by photon echo beating \textit{Laser Phys. Lett. } \textbf{13} 075202

\bibitem{6} Macfarlane R, Harris T, et. al. 1997 Measurement of photon echoes in Er:Y2SiO5 at 1.5 mm with a diode laser and an amplifier  \textit{Optics Letters} \textbf{22} 871

\bibitem{7} Maximov B, Charitonov Yu, et. al. 1968
\textit{Dokl. Akad. Nauk SSSR, Ser. Mat.
Fiz. } \textbf{183} 1072; Maksimov, B, Iljukhin V, et.al. 1990 \textit{Kristallographiya}  \textbf{35} 610 (in Russian)

\bibitem{8}  Graf F, Renn A, et. al. 1997 Site interference in Stark-modulated photon echoes \textit{Phys. Rev. B } \textbf{55} 11225

 \bibitem{9}  Lisin V, Shegeda A, Samartsev V,  Chukalina E 2018 Effect of the Excitation Radiation Coherence on Oscillations of the Photon Echo Intensity  \textit{JETP Letters} \textbf{107}
345-350

 \bibitem{10}  Doualan J, Labbe C., et. al. 1995 Energy levels of the laser active Er3+ ion in each of the two crystallographic sites of yttrium orthosilicate  \textit{Journal of Physics: Condensed Matter} \textbf{7}
5111

 \bibitem{11}  Wang Y, Meltzer R 1992 Modulation of photon-echo intensities by electric fields: Pseudo-Stark splittings
in alexandrite and YAI03:Er3+ \textit{Phys. Rev. B } \textbf{45}
10119

 \bibitem{12}  Hastings-Simon S, Staudt M, et. al. 2006 Controlled Stark shifts in Er3+ -doped crystalline and amorphous waveguides for quantum state storage \textit{arXiv:quant-ph/0603194v2}

\endbib
\end{document}